\documentclass[useAMS,usenatbib]{mn2e}
\usepackage{graphicx}
\usepackage{natbib}
\usepackage{amssymb}
\usepackage{setspace}
\usepackage{amssymb}
\usepackage{epsfig,color}
\usepackage{multicol}

\long\def\symbolfootnote[#1]#2{\begingroup%
\def\thefootnote{\fnsymbol{footnote}}\footnote[#1]{#2}\endgroup} 

\title[Structure and density distribution of MCs]{Statistical link between the structure of molecular clouds and their density distribution}

\author[Donkov, Veltchev \& Klessen]
{	
\parbox{\textwidth}{Sava Donkov$^{1\,\star}$, Todor V.~Veltchev$^{2,3}$, and Ralf S. Klessen$^3$}\vspace{0.4cm} \\
  $^1$Department of Applied Physics, Technical University, 8 Kliment Ohridski Blvd., 1000 Sofia, Bulgaria \\
  $^2$University of Sofia, Faculty of Physics, 5 James Bourchier Blvd., 1164 Sofia, Bulgaria\\
  $^3$Universit\"at Heidelberg, Zentrum f\"ur Astronomie, Institut f\"ur Theoretische Astrophysik, Albert-Ueberle-Str. 2, 69120 Heidelberg, Germany\\
}

\date{Accepted 2016 December 1. Received 2016 November 1; in original form 2016 June 21
 }

\pagerange{\pageref{firstpage}--\pageref{lastpage}} \pubyear{2016}

\begin{document}

\label{firstpage}

\maketitle

\begin{abstract}
We introduce the concept of a class of equivalence of molecular clouds represented by an abstract spherically symmetric, isotropic object. This object is described by use of abstract scales in respect to a given mass density distribution. Mass and average density are ascribed to each scale and thus are linked to the density distribution: a power-law type and an arbitrary continuous one. In the latter case, we derive a differential relationship between the mean density at a given scale and the structure parameter which defines the mass-density relationship. The two-dimensional (2D) projection of the cloud along the line of sight is also investigated. Scaling relations of mass and mean density are derived in the considered cases of power-law and arbitrary continuous distributions. We obtain relations between scaling exponents in the 2D and 3D cases. The proposed classes of equivalence are representative for the general structure of real clouds with various types of column-density distributions: power law, lognormal or combination of both.
\end{abstract}

\begin{keywords}
ISM: clouds - ISM: structure - methods: statistical
\end{keywords}

\section{Introduction}   \label{Sec-Intr}

Molecular clouds (MCs) are the birthplaces of stars in galaxies. Their morphological and kinematical evolution is governed by complex physics including gravity, supersonic turbulent flows, magnetic fields, feedback from young massive stars. MCs form from interstellar warm atomic gas %(mostly hydrogen)
% ,  is a result of colision of two or several  macro-streams (sizes about $150\times50\times50 \rm~pc$) withti temperatures $5000-10000 \rm~K$, number density $\sim 0.1 \rm~cm^{-3}$ and Mach numbers (of the turbulent velosity fluctuations) $\sim 1-2$. 
% , at the place of collision, begins cooling rapidly by isobaric contraction
which is compressed by supersonic flows and cools down rapidly due to non-linear thermal instabilities \citep{VS_ea_07, Baner_ea_09}, reaching temperatures $10-30$~K, number densities $\sim 100 \rm~cm^{-3}$ and turbulent velocity Mach numbers ${\cal M}\sim 20-30$. The gas becomes molecular due to shielding from the ambient interstellar radiation field and dense regions can be identified observationally as (giant) molecular cloud complexes. Their subsequent evolution is determined basically by supersonic compressible turbulence and self-gravity, acting simultaneously \citep{Klessen_00, Elmegreen_07, Kritsuk_ea_07, VS_10}. 

At the very begining of MC evolution (a few Myr) supersonic turbulence generates complex nets of condensations and dilutions which results in a lognormal probability distribution of mass density \citep{Klessen_00, Kritsuk_ea_07, Federrath_ea_10}, i.e., a Gaussian distribution of the logdensity. Such type of probability distribution function (hereafter, pdf) is in agreement with the observed hierarchical (fractal) structure of gas condensations in Galactic clouds \citep{Elmegreen_Falgarone_96, Elmegreen_02}. At a later evolutionary stage of $\sim10$~Myr \citep{VS_ea_07}, self-gravity becomes significant in the energy budget of the cloud. Gravitational and kinetic energy reach roughly equipartition at different spatial scales \citep{BP_VS_95, BP_06, DVK_11} and different cloud substructures start to collapse on their own free-fall timescales. The cloud undergoes a so called hierarchical gravitational collapse \citep{BP_ea_11}. Meanwhile the warm atomic gas from the cloud halo continues to accrete onto the cloud \citep{KH_10}, causing an increase of its mass. The gas gets cooler; it crosses the region of phase transition between the atomic and molecular phase. Material moves through all fractal scales under the influence of supersonic turbulence and gravity; at the lower scale limit of this gravoturbulent cascade a fraction of the gas is transformed to stars while most of it disperses again in the interstellar medium. Within a period of several Myr, before young stars emerge and provide feedback, one could consider all processes being in a rough equilibrium at all hierarchical scales and at timescales significantly smaller than the dynamical cloud time\footnote{It is about the free-fall time and the turbulent crossing time, under the assumption of equipartition between gravitational and kinetic energy.}. This equilibrium is essentially statistical and should be understood as referring to averaged quantities and to abstract objects like fractal scales. 
% (turbulence itself, supernovae explosions and radiating from massive stars cause that). 

The originally lognormal density pdf gradually changes under the influence of self-gravity. Its high-density part transforms into a power-law (PL) `tail' with a negative slope. Gas condensations, in which stars form, are representative of this density regime. The slope of the PL tail gets slowly shallower while the lognormal component of the pdf shape undergoes minor changes \citep{Girichidis_ea_14} -- another hint to a statistical equilibrium. Thus the density pdf is an important tool to study the cloud physics. It bears signature of the cloud evolution, contains information about the cloud structure and is verifiable from numerical simulations and observations. On the other hand, saturated supersonic turbulence and gravity create a self-similar hierarchy of scales which can be considered as fractal cloud structure characterized by scaling laws of mass, density and velocity fluctuation \citep{Larson_81, Schneider_ea_11, DVK_11, DVK_12, VDK_13, Girichidis_ea_14}. Establishing a link between the notion of abstract scale and the density pdf is, therefore, fundamental for development of a more systematic theory of MCs. 

In this Paper we propose a statistical model that quantifies this link. Section \ref{Sec-Statstr} introduces the basic concepts like MC class of equivalence, abstract fractal scales and their relation to the density pdf. Then, the cases of a PL pdf (Section \ref{Sec-PL-tail}) and an arbitrary continuous pdf (Section \ref{Sec-arbitrary pdf}) are considered, with derivation of the scaling relations from analysis of the pdf and of a parameter which characterizes the cloud structure. Section \ref{Sec-2D case} is dedicated to the cloud structure as described by surface mass density distribution and its relation to the 3D (mass density) case. In Section \ref{Discussion} we discuss on the applicability of the model to several general types of observational and numerical pdfs. Section \ref{Sec-Disc-Concl} contains our conclusions.

% To include in Discussion: understanding of how is built the fractal MC hierarchy. This, also, can give a link between the probability distribution of gas condensations (so called Clump mass function) and Initial stellar mass function.

% In that we see the main contribution of the paper. It might be as well to mention the differential relationship (\ref{diff-relationship-3D}) between averaged density at a given scale and structure parameter (of that scale) which allows us to calculate easy the structure parameter in the case of PL-tail.
% , which determines existing of scalling laws for turbulent velocity fluctuations, mean density and mass of stuctures and substuctures in the fracral (Kritsuk et al 2007; Larson 1981)
%Also the fractal structure and the concomitant scaling laws of the whole cloud are conserved. 

% \begin{figure*} 
% \begin{center}
% \includegraphics[width=.85\textwidth]{fig_pdf.eps}
% \vspace{0.6cm}
% \caption{Probability distribution function (pdf) of mass density: (left) lognormal, from purely turbulent simulation \citep{Kritsuk_ea_07}; (right) a combination of lognormal and PL-tail, from simulation with gravity \citep[][the slopes for volume- and mass-weighted pdf are given]{Girichidis_ea_14}. }
% Both axes are built in logarithmic scale - $s=\ln(\rho/ \langle\rho \rangle _{\rm c})$ on abcice and $\ln(p(s))$ on ordinat, where $p(s)=B\exp(-(s-\bar{s})^2/2\sigma^2)$ or $p(s)=A\exp(qs)$.}
% \label{fig_pdfs_simulations}
% \end{center}
% \end{figure*}

\section{The statistical MC structure}   
\label{Sec-Statstr}

Let us consider -- for the purpose of our general study, -- an abstract model of MC. The cloud is spherical with mass density profile $\rho(\ell)$, determined from the mass density pdf in the real medium. The scales $\ell$ in that model are defined simply as radii measured from the centre of the sphere to a given density level and span the range $l_{0} \ll \ell \leq l_{\rm c}$, where $l_{\rm c}$ is the size of the whole cloud and $l_{0}$ is the size of its homogeneous core. 

The proposed model is representative of a {\it MC class of equivalence}. By assumption, all class members are characterized by single mass density pdf, single cloud size, single size, density of the core and density at the cloud's edge. We stress that individual members could have very different morphology and, probably, different physics, but for the purpose of this work the abovementioned characteristics are sufficient to describe the class. In other words, an MC class of equivalence resembles a statistical ensemble for which the averaged member possesses spherical symmetry and isotropy. This averaged member is an abstract object which is statistically representative for the behaviour of any single member of the class. It is conceptually similar to the spherically symmetric nested MC model of \citet{LB_16}; the difference is that those authors use it as a simplification of the real cloud.

In the following we will use mainly the volume-weighted mass density pdf $p(s)$. The natural variable in the model is the logarithmic density $s=\ln(\rho/\langle\rho\rangle_{\rm c})$, where $\langle\rho\rangle_{\rm c}$ is the average density of the whole cloud. The abstract scales of the fractal are defined as follows:
\begin{equation}
\label{def_abstr-scale}
\ell(s)=l_{\rm c}\Big(\int\limits_{s}^{\infty}p(s)ds\Big)^{1/3}~,
\end{equation}
where the upper integration limit is taken to be infinity, i.e. it corresponds to very large densities. Thus $\ell(s)$ is radius of the sphere, corresponding to density level $\rho=\langle\rho\rangle_{\rm c}\exp(s)$. The suggested definition of scale is abstract in the sense that $\ell(s)$ is not related to size of any contiguous objects, delineated by analysis of MC intensity maps or through some clump extraction techniques. On the other hand, it is not a purely mathematical construct like, e.g. the lag in the $\Delta$-variance method \citep{Stutzki_ea_98} or the quantity defining the wave number in Fourier analysis of star-forming regions. It contains implicitly the physics of the considered MC through the mass density pdf. Similar definition is adopted by \citet{LB_16} but on an additional assumption, simplifying the cloud structure.

%%% To clarify the relation between abstract scale, defined through the pdf, and scales, defined/measured on the basis of some observational/numerical approach
In the definition of scale (equation \ref{def_abstr-scale}) we neglect the size of the core $l_0 \ll \ell$ to simplify the calculations. This generates a singularity in the expression for density profile (equation \ref{dens_profile}) which is not significant in the further considerations and does not affect the results.

\section{Power-law density pdf}   \label{Sec-PL-tail}

\subsection{Relation to the density profile}   
\label{subsec-PL-tail_profile}

Let us introduce a power-law density pdf:
\begin{equation}
\label{def_PL-tail}
p(s)=A\exp(qs)=A\exp(q\ln(\rho/\rho_{\rm c}))=A(\rho/\rho_{\rm c})^q~,
\end{equation}
where the dimensionless exponent typically spans the range $-3\leq q \leq-1.5$ \citep{BP_ea_12, FK_13, Girichidis_ea_14}. For the sake of convenience in calculations, we redefine $s=\ln(\rho/\rho_{\rm c})$, where $\rho_{\rm c}$ is the density at cloud's outer edge. As will be demonstrated later, $\rho_{\rm c}$ and $\langle\rho\rangle_{\rm c}$ are equivalent within a factor of few. It is worth to note that $\rho_{0}\gg\rho(s)\geq\rho_{\rm c}$, where $\rho_{0}$ is the core density. The constant $A$ in equation \ref{def_PL-tail} can be derived from the normalization condition of probability:

\begin{equation}
\label{norm_pdf}
\int\limits_{\rho_{\rm c}}^{\rho_{0}}A\exp(qs)ds=1~~\Rightarrow~~A=q\Big/\big[(\rho_{0}/\rho_{\rm c})^q-1\big]\simeq -q
\end{equation}

Per definition, the volume-weighted pdf is:

\begin{equation}
\label{def_pdf}
p(s)\equiv \frac{d\,V(s)/\Delta V}{d\ln(\rho(s)/\rho_{\rm c})}~,
\end{equation}
where $V(s)\equiv 4\pi \ell^3(s)/3$ is the volume of a sphere with radius $\ell(s)$ and $\Delta V=V_{\rm c}-V_{0}=4\pi(l_{\rm c}^3-l_{0}^3)/3$ is the difference between cloud volume and core volume. This follows from the normalization:

$$\int\limits_{\rho_{\rm c}}^{\rho_{0}}d(V(s)/\Delta V)=1~.$$

Now let's consider the differential equation $d(V(s)/\Delta V)=A\exp(qs)ds$ and integrate it in the range $[\rho_{\rm c}, \rho_{0}]$. Then we obtain:

$$\frac{V(s)-V_{0}}{V_{\rm c}-V_{0}}=\frac{A}{q}\Big[\Big(\frac{\rho_{0}}{\rho_{\rm c}}\Big)^q-\Big(\frac{\rho(s)}{\rho_{\rm c}}\Big)^q\Big]$$

In view of the relations $\rho_{0}\gg\rho(s)\geq\rho_{\rm c}$ and $l_{0}\ll \ell(s)\leq l_{\rm c}$, one gets $V(s)/V_{\rm c}=\ell^3(s)/l^3_{\rm c}$ for the l.h.s. and $(\rho(s)/\rho_{\rm c})^q$ for the r.h.s., accordingly, and hence the density profile in the case of a PL pdf\footnote{For the sake of completeness of these considerations, let us note that from a density profile in the form (\ref{dens_profile}), one can derive -- according to (\ref{def_pdf}) -- the same PL pdf like in (\ref{def_PL-tail}). In other words, there is an unambiguous correspondence between a density profile with constant exponent and a PL pdf \citep{KNW_11, Girichidis_ea_14}.}:

\begin{equation}
\label{dens_profile}
\rho(\ell(s))\simeq \rho_{\rm c}\Big(\frac{\ell(s)}{l_{\rm c}}\Big)^\frac{3}{q}~.
\end{equation}

The exponent in this equation is usually denoted $-p$ in the literature, where $p$ is positive and labelled ``density profile''. Clearly,
\[ p=-3/q~~\Leftrightarrow~~q=-3/p~. \]

\subsection{Scaling of mass and averaged density}
\label{subsec-PL-tail_mass_av-density}

Consider now to the connection between a PL pdf with the scaling relations of mass and averaged density. The latter can be written in the form:

\begin{equation}
\label{M_L and rho_L}
M(\ell)= M_{\rm c} \Big(\frac{\ell}{l_{\rm c}}\Big)^{\gamma}~~,~~\langle\rho\rangle_{\rm \ell}\equiv\frac{M(\ell)}{V(\ell)}=\langle\rho\rangle_{\rm c} \Big(\frac{\ell}{l_{\rm c}}\Big)^{\alpha}~,
\end{equation}

where $\ell=\ell(s)$ is the considered scale and $M_{\rm c}$ is the mass of the whole cloud. Following the considerations from the previous Section, we aim to express the scaling exponents $\gamma$ and $\alpha$ through the PL slope $q$. If $M_{0}=\rho_{0}\,4\pi l^3_{0}/3$ is the core mass, one can calculate the mass $M(\ell)$ of a given scale from equation (\ref{dens_profile}):
\begin{eqnarray}
\label{M_L-calc}
M(\ell) & = & \int\limits_{l_{0}}^{\ell}4\pi\rho(\ell)\ell^2d\ell + M_{0} \nonumber \\ 
~ & = & 4\pi\rho_{\rm c}l^3_{\rm c}\int\limits_{l_{0}}^{\ell}(\ell/l_{\rm c})^{2-p}d(\ell/l_{\rm c}) + M_{0} \nonumber \\ 
~ & = & \frac{4\pi}{3-p}\rho_{\rm c}l^3_{\rm c}\big[(\ell/l_{\rm c})^{3-p} - (l_{0}/l_{\rm c})^{3-p}\big] + M_{0} \nonumber \\ 
~ & \simeq & \frac{4\pi}{3-p}\rho_{\rm c}l^3_{\rm c}\,\Big(\frac{\ell}{l_{\rm c}}\Big)^{3-p}~,
\end{eqnarray}

where we neglect the second addend in the parentheses and the core mass $M_{0}$, taking into account that $1\leq p \leq2$, $(l_{0}/\ell)\ll1$ and thus $(l_{0}/\ell)^{3-p}\ll1$ as well:

\[\frac{M_0}{\frac{4\pi}{3-p}\rho_{\rm c}l^3_{\rm c}(\ell/l_{\rm c})^{3-p}}=\frac{3-p}{3}\Big(\frac{l_{0}}{\ell}\Big)^{3-p}\ll1~.\]

Setting $\ell=l_{\rm c}$ and  comparing (\ref{M_L-calc}) and (\ref{M_L and rho_L}) leads to:
\[ M_{\rm c}=\frac{4\pi}{3-p}\,\rho_{\rm c}l^3_{\rm c}~,~~~\gamma=3-p=3+\frac{3}{q} \] 
\[ \langle\rho\rangle_{\rm c}=\frac{3}{3-p}\,\rho_{\rm c}~,~~~\alpha=-p=\frac{3}{q} \]

\subsection{Parameter of cloud structure and its relation to the slope $q$}
\label{subsec-str-parameter_PL-tail}

The relationship between mass and averaged density of hierarchical structures in the cloud is essential to demonstrate a link between the cloud's fractal morphology as characterized by the pdf and the corresponding mass function of these structures. Let's define a power-law relationship between mass and averaged density of given scale:

\begin{equation}
\label{rho-m_relationship}
\frac{\langle\rho\rangle_{\rm \ell}}{\langle\rho\rangle_{\rm c}}=\Big(\frac{M_{\rm \ell}}{M_{\rm c}}\Big)^{x}~,
\end{equation}

where $x$ is labelled ``parameter of cloud structure'' or, simply, {\it structure parameter} \citep[see][]{VKC_11, DVK_11}. From equation (\ref{M_L and rho_L}) and using the relations: $\gamma=3-p$, $\alpha=-p$, $p=-3/q$, we derive a relation between the indices $x$ and $q$: $x=1/(1+q)$. There is an evident, but important, relationship between the normalized mass, the averaged density and the volume (size) of a given scale:

\[ \frac{M(\ell)}{M_{\rm c}} = \Big(\frac{\ell}{l_{\rm c}}\Big)^{\gamma}=\Big(\frac{\ell}{l_{\rm c}}\Big)^{3-p}  =\frac{\langle\rho\rangle_{\rm \ell}}{\langle\rho\rangle_{\rm c}}\Big(\frac{\ell}{l_{\rm c}}\Big)^{3}~\Rightarrow  \]
\begin{equation}
\label{norm-relation_PL-tail}
\frac{M(\ell)}{M_{\rm c}}=\frac{\langle\rho\rangle_{\rm \ell}}{\langle\rho\rangle_{\rm c}}\frac{V(\ell)}{V_{\rm c}}
\end{equation}

Generally, the density profile $p$ and the scaling indices $\gamma$ (of the mass), $\alpha$ (of the averaged density) and $x$ (structure parameter) are expressed through the PL slope $q$ as follows:

\begin{equation}
\label{exponents through q}
p=-\frac{3}{q},~~\gamma=3\frac{1+q}{q},~~\alpha=\frac{3}{q},~~x=\frac{1}{1+q}
\end{equation}

The usefulness of these relations becomes evident in Sect. \ref{subsec-2D case-scales and scaling laws} where we derive a relation between the PL slope $q$ of the density pdf and the observable PL slope of the column-density pdf. Hence one gets a link to the general cloud structure as described through the structure parameter $x$ and, further, to the mass function of hierarchical structures or even to the initial stellar mass function \citep{DVK_12, VDK_13}.

\section{Arbitrary pdf}	
\label{Sec-arbitrary pdf}

\subsection{Scales and scaling relations}
\label{subsec-scales and scaling laws}

Now, recalling the definition of log-density $s=\ln(\rho/\langle\rho\rangle_{\rm c})$ (Section \ref{Sec-Statstr}), we consider a volume-weighted normalized pdf $p(s)$ of {\it arbitrary shape} with the only requirement of continuity:

\begin{equation}
\label{pdf-norm condition}
\int\limits_{-\infty}^{\infty} p(s)ds=1~.
\end{equation}

Accordingly, mass, volume and averaged density of cloud structures above some density cut-off level $s'$ and their derivatives are defined:

\begin{eqnarray}
\label{V-dV}
\frac{V'}{V_{\rm c}}=\int\limits_{s'}^{\infty} p(s)ds~~\Rightarrow~~\frac{d}{ds'}\Big(\frac{V'}{V_{\rm c}}\Big)= -p(s')~,
\end{eqnarray}

\begin{eqnarray}
\label{M-dM}
\frac{M'}{M_{\rm c}}=\!\int\limits_{s'}^{\infty}\exp(s) p(s)ds~\Rightarrow
\frac{d}{ds'}\Big(\frac{M'}{M_{\rm c}}\Big)\!\!= -\exp(s')p(s')~.
\end{eqnarray}

Combining both formulae, one obtains:

\begin{equation}
\label{dM-dV}
\frac{d}{ds'}\Big(\frac{M'}{M_{\rm c}}\Big)=
\exp(s')\frac{d}{ds'}\Big(\frac{V'}{V_{\rm c}}\Big)
\end{equation}

and

\begin{eqnarray}
\label{avrho_pr-norm_relationship}
\rho'=\frac{M'}{V'}=\langle\rho\rangle_{\rm c} \int\limits_{s'}^{\infty}\exp(s)p(s)ds \Big/\int\limits_{s'}^{\infty}p(s)ds~~\Rightarrow
\end{eqnarray}

\begin{equation}
\label{norm_relationship_mass_rho_V}
\frac{M'}{M_{\rm c}}=\frac{\rho'}{\langle\rho\rangle_{\rm c}}\frac{V'}{V_{\rm c}}~.
\end{equation}

A generalized notion of scale is introduced with one-to-one correspondence to the density cut-off level:

\begin{equation}
\label{L_pr}
\ell'(s')=l_{\rm c}\Big(\int\limits_{s'}^{\infty}p(s)ds\Big)^{1/3}~.
\end{equation}

Hence one derives the scaling relations of mass and averaged density with indices, depending on $s'$:

\begin{equation}
\label{gen_scaling-laws}
\frac{M'}{M_{\rm c}}=\Big(\frac{\ell'}{l_{\rm c}}\Big)^{\gamma(s')}~~,~~\frac{\rho'}{\langle\rho\rangle_{\rm c}}=\Big(\frac{\ell'}{l_{\rm c}}\Big)^{\alpha(s')}~,
\end{equation}

where the normalizing quantity is the effective radius of the cloud: $V_{\rm c}=\frac{4}{3}\pi l^{3}_{\rm c}$. 

Assuming a power-law mass-density relation of type

\begin{eqnarray}
\label{mass-density_general}
\frac{\rho'}{\langle\rho\rangle_{\rm c}}=\Big(\frac{M'}{M_{\rm c}}\Big)^{x(s')}
\end{eqnarray}

the scaling indices of mass and density are interdependent in view of equations (\ref{norm_relationship_mass_rho_V}) and (\ref{gen_scaling-laws}):

\begin{equation}
\label{gen_scal-exp_relation}
\gamma=3+\alpha~,~~\gamma x=\alpha~.
\end{equation}

Thus the knowledge of $x$ yields $\alpha$ and $\gamma$ from (\ref{gen_scal-exp_relation}). On the other hand, the exponent $x$ can be estimated by use of (\ref{M-dM}), (\ref{avrho_pr-norm_relationship}) and (\ref{mass-density_general}):

\begin{equation}
\label{gen_x}
x(s')=1-\ln\Big(\int\limits_{s'}^{\infty} p(s)ds\Big)\Big/\ln\Big(\int\limits_{s'}^{\infty} \exp(s)p(s)ds\Big)
\end{equation}

Hence the knowledge of the pdf $p(s)$ determines -- according to our model -- the cloud structure in terms of abstract scales and scaling relations with indices, which are functions of the cut-off level $s'$. Also, one can derive the mass function of hierarchical cloud structures using (\ref{mass-density_general}).

\subsection{Differential relationship between the structure parameter and the averaged scale density}
\label{subsec-Diff-relationship}

To derive a differential relationship between the general structure parameter $x(s')$ and the averaged scale density at a given cut-off level, we rewrite the normalized relationship between scale mass, averaged density and volume (equation \ref{norm_relationship_mass_rho_V}) in the form

$$\frac{V'}{V_{\rm c}}=\Big(\frac{\rho'}{\langle\rho\rangle_{\rm c}}\Big)^{-1}\frac{M'}{M_{\rm c}}$$

and take the derivative in respect to $s'$:

$$\frac{d}{ds'}\Big(\frac{V'}{V_{\rm c}}\Big)=\Big({\!}\frac{\rho'}{\langle\rho\rangle_{\rm c}}{\!}\Big)^{-1}\!\!\frac{d}{ds'}{\!}\Big(\frac{M'}{M_{\rm c}}\Big)-\Big({\!}\frac{\rho'}{\langle\rho\rangle_{\rm c}}{\!}\Big)^{-2}{\!}\frac{M'}{M_{\rm c}}\frac{d}{ds'}{\!}\Big({\!}\frac{\rho'}{\langle\rho\rangle_{\rm c}}\Big)$$

Making use of equation \ref{dM-dV}, one gets after several simple algebraic transformations:

\begin{equation}
\label{midle equ}
\frac{d}{ds'}\ln\Big(\frac{\rho'}{\langle\rho\rangle_{\rm c}}\Big)=\Big[1-\exp(-s')\Big(\frac{\rho'}{\langle\rho\rangle_{\rm c}}\Big)\Big]\frac{d}{ds'}\ln\Big(\frac{M'}{M_{\rm c}}\Big)~.
\end{equation}

On the other hand, the differentiation of the logarithmic relationship between scale mass and density (equation \ref{mass-density_general}) yields:

\begin{equation}
\label{diff_mass-density_general}
\frac{d}{ds'}\ln\Big(\frac{\rho'}{\langle\rho\rangle_{\rm c}}\Big)=x\frac{d}{ds'}\ln\Big(\frac{M'}{M_{\rm c}}\Big)+\ln\Big(\frac{M'}{M_{\rm c}}\Big)\frac{d}{ds'}x
\end{equation}

Now, we substitute $d\ln(M'/M_{\rm c})/ds'$ from equation (\ref{midle equ}) and $\ln(M'/M_{\rm c})$ from the mass-density relation (equation \ref{mass-density_general}) in the formula (\ref{diff_mass-density_general}) above. Lastly, after some further algebraic transformations, we obtain:

\begin{eqnarray}
\label{diff-relationship-3D}
\Big[1-\exp(-s')\Big(\frac{\rho'}{\langle\rho\rangle_{\rm c}}\Big)-x\Big]\frac{d}{ds'}\ln\Big(\frac{\rho'}{\langle\rho\rangle_{\rm c}}\Big)= \nonumber \\
\ln\Big(\frac{\rho'}{\langle\rho\rangle_{\rm c}}\Big)\Big[1-\exp(-s')\Big(\frac{\rho'}{\langle\rho\rangle_{\rm c}}\Big)\Big]\frac{d}{ds'}\ln(x)~.
\end{eqnarray}

Note that when an arbitrary pdf $p(s)$ is given and the formula (\ref{diff-relationship-3D}) is considered as a differential equation for $x(s')$, the latter has an exact solution, namely expression (\ref{gen_x}).

The case of PL-type pdf yields an interesting application of equation (\ref{diff-relationship-3D}): its r.h.s. is equal to zero since $d\ln(x)/ds'=0$. Then the l.h.s. is also zero which is only possible if the expression in the parentheses is zero. (Since $d\ln(\rho'/\langle\rho\rangle_{\rm c})/ds'\neq 0$ for every non-homogeneous cloud.) Hence we get an equation for $x$:

\begin{eqnarray}
\label{equ_x-diff}
x=1-\exp(-s')\Big(\frac{\rho'}{\langle\rho\rangle_{\rm c}}\Big)= \nonumber \\ 
1-\exp(-s')\int\limits_{s'}^{\infty} \exp(s)p(s)ds\Big/\int\limits_{s'}^{\infty} p(s)ds~.
\end{eqnarray}

From the explicit expression for PL-pdf (equation \ref{def_PL-tail}), one obtains $x=1/(1+q)$. This is an independent confirmation of equation (\ref{exponents through q}). We point out that the differential relationship (\ref{diff-relationship-3D}) is derived with the only restriction for continuity of the pdf. This gives opportunities to derive the structure parameter for wide variety of pdf form.

\section{The 2D case under the condition of point symmetry}
\label{Sec-2D case}

\subsection{Scales and scaling laws in the 2D case} \label{subsec-2D case-scales and scaling laws}

Let $\Sigma$ be the surface mass density of the observed object (cloud) and  $M_{\rm c}$ and $S_{\rm c}$ are its mass and projected surface area, respectively. Then $\langle\Sigma\rangle_{\rm c}=M_{\rm c}/S_{\rm c}$ is the averaged surface mass density of the cloud and $z=\ln(\Sigma/\langle\Sigma\rangle_{\rm c})$ is the logarithmic surface mass density.

Let $p(z)$ be the probability distribution function of the surface mass density, labelled pdf for simplicity, with the only restriction for continuity. For a fixed cut-off level $z'=\ln(\Sigma'/\langle\Sigma\rangle_{\rm c})$ of the logarithmic surface mass density, one could introduce the corresponding mass, projected surface area, scale and averaged surface mass density:

\begin{eqnarray}
\label{M_S_L-2D}
M'=M_{\rm c}\int\limits_{z'}^{\infty} \exp(z)p(z)dz \nonumber \\ 
S'=S_{\rm c}\int\limits_{z'}^{\infty} p(z)dz \nonumber \\ 
\ell'(z')=l_{\rm c}\Big(\int\limits_{z'}^{\infty} p(z)dz\Big)^{1/2}~~,~~l_{\rm c}=\sqrt{S_{\rm c}/\pi} \nonumber \\ 
\Sigma'=\frac{M'}{S'}=\langle\Sigma\rangle_{\rm c}\int\limits_{z'}^{\infty} \exp(z)p(z)dz\,\Big/ \int\limits_{z'}^{\infty} p(z)dz~.
\end{eqnarray}

Like in the 3D case (equation \ref{norm_relationship_mass_rho_V}), there is a relationship between normalized mass, averaged (surface) density and projected surface area:

\begin{equation}
\label{norm relationship-2D}
\frac{M'}{M_{\rm c}}=\frac{\Sigma'}{\langle\Sigma\rangle_{\rm c}}\frac{S'}{S_{\rm c}}
\end{equation}

Here we introduce a relationship between mass and averaged surface density:

\begin{equation}
\label{mass density-2D}
\frac{\Sigma'}{\langle\Sigma\rangle_{\rm c}}=\Big(\frac{M'}{M_{\rm c}}\Big)^{y(z')}~,
\end{equation}

where $y$ is called {\it structure parameter} in the 2D case.

\subsection{The relations between scaling indices in the 2D and 3D cases}
\label{Relations scaling exponents 2D-3D}

The key assumption is that the cloud possesses point symmetry which stems from our statistical model and reads:

\begin{equation}
\label{isotropic equ}
\rho'\propto\Sigma'(\ell')^{-1}~,
\end{equation}

and, in normalized form,

\begin{equation}
\label{isotropic equ-norn}
\frac{\rho'}{\langle\rho\rangle_{\rm c}}=\frac{\Sigma'}{\langle\Sigma\rangle_{\rm c}}\Big(\frac{\ell'}{l_{\rm c}}\Big)^{-1}~.
\end{equation}

Now from equations (\ref{rho-m_relationship}), (\ref{gen_scaling-laws}), (\ref{gen_scal-exp_relation}) and (\ref{mass density-2D}) one could obtain after several algebraic transformations : $(M'/M_{\rm c})^{x}=(M'/M_{\rm c})^{[y-(1-x)/3]}$. This leads to relations between the scaling indices in the 2D and the 3D case:

\begin{eqnarray}
\label{connect-scale exp_2D-3D}
y=\frac{2x+1}{3}~~\Leftrightarrow~~x=\frac{3y-1}{2} \nonumber \\ 
y=\frac{\gamma-2}{\gamma}~~\Leftrightarrow~~\gamma=\frac{2}{1-y}~.
\end{eqnarray}

The relations above allow for writing the scaling relations for mass and averaged surface density in the 2D case:

\begin{equation}
\label{gen_scaling-laws_2D}
\frac{M'}{M_{\rm c}}=\Big(\frac{\ell'}{l_{\rm c}}\Big)^{\gamma}~~,~~\frac{\Sigma'}{\langle\Sigma\rangle_{\rm c}}=\Big(\frac{\ell'}{l_{\rm c}}\Big)^{\gamma-2}~.
\end{equation}

An interesting special case is the pdf of PL type with constant $x(s')$. As seen in Section \ref{Sec-PL-tail}, in a spherically symmetric cloud with pdf $p(s)=A\exp(qs)$ one could obtain a PL mass density profile $\rho(\ell)\propto \ell^{-p}$ where $q=-3/p$. Here we consider a 2D object which is a projection of the cloud along the line of sight. This projection is simply a circle and the pdf of the surface mass density is also of PL type: $p(z)=C\exp(nz)$. The latter corresponds to a surface mass density profile $\Sigma(\ell)\propto \ell^{-t}$ where $n=-2/t$. Note that there is also a relation between two density-profile indices $t=p-1$ and, hence, $2/n=3/q+1$. These relations between indices for a PL pdf in the 3D and 2D cases can be found elsewhere in the literature \citep[e.g.][]{KNW_11, BP_ea_12,  FK_13, Girichidis_ea_14}. 

Also, we point out two useful relations in the case of PL-pdf:

\begin{equation}
\label{y-gamma-n-q_relation}
y=\frac{1}{1+n}=\frac{3+q}{3(1+q)}~~,~~\gamma=2\frac{1+n}{n}=3\frac{1+q}{q}~.
\end{equation}

They can be derived in the same way like in the 3D consideration (Section \ref{Sec-PL-tail}).
%%% Constraints (observational and numerical) on the estimates of the x parameter and hence -- through the proposed model -- on the slope of the CMF/IMF. Comments.

\subsection{Differential relationship between the structure parameter and the averaged surface mass density} \label{subsec-diff relation in 2D}

Finally, we present briefly the differential relationship between the structure parameter in the 2D case $y$ and the averaged surface density at a given cut-off level $z'=\ln(\Sigma'/\langle\Sigma\rangle_{\rm c})$. It is easy to see from equations (\ref{M_S_L-2D}) that

$$\frac{d}{dz'}\Big(\frac{S'}{S_{\rm c}}\Big)=-p(z')~,~\frac{d}{dz'}\Big(\frac{M'}{M_{\rm c}}\Big)=-\exp(z')p(z')$$
$$\Rightarrow~~\frac{d}{dz'}\Big(\frac{M'}{M_{\rm c}}\Big)=\exp(z')\frac{d}{dz'}\Big(\frac{S'}{S_{\rm c}}\Big)~.$$

From identical transformations like in Subsection \ref{subsec-Diff-relationship} we obtain the corresponding differential relationship in the 2D case:

\begin{eqnarray}
\label{diff-relationship-2D}
\Big[1-\exp(-z')\Big(\frac{\Sigma'}{\langle\Sigma\rangle_{\rm c}}\Big)-y\Big]\frac{d}{dz'}\ln\Big(\frac{\Sigma'}{\langle\Sigma\rangle_{\rm c}}\Big)= \nonumber \\ 
\ln\Big(\frac{\Sigma'}{\langle\Sigma\rangle_{\rm c}}\Big)\Big[1-\exp(-z')\Big(\frac{\Sigma'}{\langle\Sigma\rangle_{\rm c}}\Big)\Big]\frac{d}{dz'}\ln(y)~.
\end{eqnarray}

The exact solution of the above equation considered for $y$ stems directly from the relation \ref{mass density-2D} and reads:

\begin{equation}
\label{gen_y}
y=1-\ln\Big(\int\limits_{z'}^{\infty} p(z)dz\Big)\Big/\ln\Big(\int\limits_{z'}^{\infty} \exp(z)p(z)dz\Big)~.
\end{equation}

Analogously to Sect. \ref{subsec-Diff-relationship}, one gets a direct application of the differential relationship (equation \ref{diff-relationship-2D}) in the case of PL-pdf. Then $d\ln(y)/dz'=0$ which sets to zero the expression in the parentheses at the l.h.s which yields an equation for $y$, very similar to  equation (\ref{equ_x-diff}), which leads to $y=1/(1+n)$. The obtained expression is identical to the first relation in the expression (\ref{y-gamma-n-q_relation}), which we derive in a different way. It is useful for study of star-forming clouds (with PL-pdfs; e.g. \citealt{Schneider_ea_15}) since it allows for investigation of their structure through simple integrations over the pdf. 
%%% Link to the constraints on the x parameter

\section{Discussion on applicability}
\label{Discussion}
The derived general formula for scaling of the structure parameter $x$ (equation \ref{gen_x}) allows to assess how our concept of the class of equivalence could be applied to various types of MCs. Below we review the most common cases from observations. Recalling the adopted condition for point symmetry of the cloud (cf. Sect. \ref{Sec-2D case}), we assume that the general shape of its volume-density pdf -- power-law, lognormal or a combination of those two forms -- is the same in the 2D case.

\subsection{Power-law density pdf}
\label{Power-law density pdf}
Though extreme, this case might be an acceptable representation for some clouds wherein the transition to the power-law part is close to the CO self-shielding limit and thus the shape of the pdf at lower column-densities is not known with certainty \citep{LAL_15}. Then the density pdf is characterized solely by the slope $q$ and the structure parameter is simple function of the latter: $x=1/(1+q)$ (equation \ref{exponents through q}) which could be reproduced also from equation (\ref{gen_x}), for plausible slopes $-3\leq q \leq-1.5$ and letting $s' \gg 1$. On the other hand, the slope $n$ of the column-density pdf and $q$ are also interrelated (Sect. \ref{Relations scaling exponents 2D-3D}): $2/n=3/q +1$ and, hence,
\begin{equation}
 \label{eq_x-n_relation_PL}
 x=(2-n)/(2+2n)~~. 
\end{equation}

Column-density pdfs of recently observed regions with some star-forming activity show well-developed PL tails, containing significant fraction of the dense gas. \citet{Abreu-Vicente_ea_15} derived average slopes between $n\sim-2$ for H {\sc ii} regions and $n\gtrsim-4$ for clouds at earlier stages of star formation. Also, \citet{Schneider_ea_15} obtained an average value $n\simeq-2.5 \pm 0.5$ from {\it Herschel} data on low- and high-mass star forming regions. Similar value of $n=-2.6$ ($q=-1.7$) results from simulations of self-gravitating isothermal and supersonic turbulent cloud at nearly half of the free-fall time \citep{KNW_11}. These findings point to typical values of the structure parameter $-2 \le x \le -1.5$ in star-forming cloud where $x=-2$ should be considered as the lower limit, which corresponds to extended PL tail with slope $q=-1.5$ in the 3D  \citep{Girichidis_ea_14} and $n=-2$ in the 2D case.

Knowledge of the structure parameter has implications for modelling the mass function of prestellar cores. If the latter form in high-density cloud parts, represented by power-law pdf and are gravitationally unstable (i.e. possible progenitors of stars), their time-averaged mass distribution should be also a power law, with slope $\Gamma=-1+x/2$, as demonstrated by \citet{DSV_12}. In view of the estimates of $x$ mentioned above, $\Gamma$ in star-forming regions should be about $-1.8$ or steeper, with lower limit $-2$ in strongly self-gravitating clouds. These slopes are steeper than the standard Salpeter value $-1.3$ of the stellar initial mass function \citep{Salpeter_55} but still within the range of observational variations \citep{Elmegreen_09}. 

\begin{figure} 
\begin{center}
\includegraphics[width=83mm]{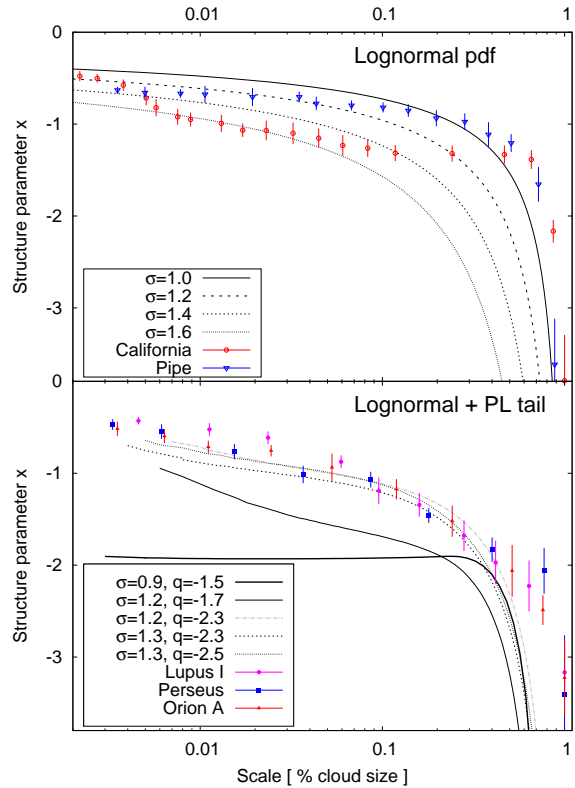}
\vspace{0.7cm}  
\caption{Scaling of the 3D structure parameter for purely lognormal pdf (top) and combined pdf (lognormal + PL tail; bottom). Observational estimates for some Galactic MCs from \citet{LAL_10} are plotted for comparison.}
\label{fig_x_L_panel}
\end{center}
\end{figure}

\subsection{Lognormal density pdf}
MCs without indications of star formation display usually a lognormal column-density distribution \citep[e.g.][]{Kainulainen_ea_09}. Recently \citet{KFH_14} proposed an approach to derive the volume-density pdf from column-density data and show that, for a number of MCs, it can be fitted by a lognormal function -- possibly excluding a small fraction of dense star-forming gas from consideration. The obtained widths span the range $1.3\lesssim \sigma \lesssim 2.1$ as the distribution peaks are weakly correlated with them. In Fig. \ref{fig_x_L_panel} (top) we plot the structure parameter vs. normalized size $\ell'/l_{\rm c}$ (cf. equation \ref{L_pr}) when $p(s)$ is a lognormal function whose peak is calculated from the width ($-\sigma^2/2$; see \citealt{VS_94}). The original observational data are obtained from extinction maps \citep{LAL_10}; then $x$ is calculated as described in \citet{DVK_11}. Both chosen clouds have lognormal column-density pdf \citep{Kainulainen_ea_09}. The shapes of the curves appear to be generally consistent with the cloud structure  except in the outer cloud parts characterized by low column-density and possibly by transition from molecular to atomic gas phase.

\subsection{Density pdf consisting of a lognormal part and a PL tail}
Most of the star-forming regions display a column-density pdf which can be decomposed to a lognormal part and a PL tail. Considered in the 3D case, such function is mathematically determined by 4 parameters: width $\sigma$ of the lognormal part, deviation point (DP), slope $q$ and density range $\Delta s$ of the PL tail. As evident from numerical and analytical studies, at earlier evolutionary stages of self-gravitating clouds the density at the deviation point is at least one order of magnitude larger than the peak density \citep{KNW_11, Collins_ea_12, Girichidis_ea_14}. For simplicity, we choose it it to vary around this value, weakly increasing with $\sigma$. Volume-density ranges were set depending on the slope $q$ which is indicative of the evolutionary stage -- well developed PL tails with $q=-1.5$ have typically $\Delta s\sim 11-14$ while $\Delta s\sim3-4$ for unevolved tails ($q<-2.$). Under these constraints, the scaling of the structure parameter is mainly a function of the lognormal width and the PL-tail slope.

In Fig. \ref{fig_x_L_panel} (bottom) is plotted the structure parameter vs. normalized size for some fiducial values of $\sigma$ and $q$. The referred observational data are for 3 clouds whose column-density pdf has a pronounced PL tail. A good agreement is achieved for unevolved PL tails and at small to intermediate scales. Well developed PL tails with $q\sim-1.5$ and large $\Delta s$ yield approximately constant $x\simeq-2$ within the cloud. This is a transition to the case of power-law column-density pdf (Sect. \ref{Power-law density pdf}) with slope $n=-2$, in full consistency with equations (\ref{exponents through q}) and (\ref{eq_x-n_relation_PL}).

\section{Conclusions and summary}
\label{Sec-Disc-Concl}
% The density probability distribution function (pdf) contains rich information about general structure of molecular clouds (MCs) and is an important signature of their physics. The MCs themselves can be considered as gas fractals since saturated supersonic turbulence and gravity generate a self-similar hierarchy of scales in them (Larson-1981, Girichidis et al-2014, Schneider et al-2014).

In this work we propose an approach to describe the general structure of molecular clouds (MCs) through a statistical object, labelled `class of equivalence'. This novel notion allows one to study a set of clouds (possibly with different morphology and physics), characterized by a single probability distribution function (pdf) of density, single total size, single size of the dense cloud core, density of the core and density at the cloud's edge. Those general parameters are considered as determining the fractal cloud structure in terms of abstract scales $\ell$, with one-to-one correspondence to given density cut-off levels $s^\prime$. In view of the scale definition (equation \ref{def_abstr-scale}) the spherical symmetry is not a simplification but a natural feature of the class representative.

The presented framework provides an useful tool for an unified investigation of the fractal structure of all class members. Scaling indices of mass $M_\ell\propto \ell^\gamma$ and averaged density $\langle \rho \rangle_\ell \propto \ell^\alpha$ are derived on very general assumptions about the density distribution (Section \ref{Sec-arbitrary pdf}). Moreover, the introduced structure parameter $x(s^\prime)$ (equation \ref{mass-density_general}) and its explicit form (equation \ref{gen_x}) allow to link the mass function of hierarchical cloud structures with the volume- and column-density pdfs \citep{DVK_11, VKC_11}. Perhaps the most important application of this scheme is the case of a power-law pdf (Section \ref{Sec-PL-tail}), studied by some other authors but in a different context \citep{KNW_11, Girichidis_ea_14, BP_ea_12}. In our treatment, it is considered as a specific sub-case of the common framework (see equation \ref{equ_x-diff}).
% (Kritsuk ea-2010). 

Our main results could be summarized as follows:
\begin{enumerate}
 \item In the general case of an arbitrary continuous pdf, the structure parameter $x=x(s^\prime)$ is an integral function of the density pdf (equation \ref{gen_x}) and determines the scaling indices of mass ($\gamma$) and averaged density ($\alpha$):
\[ \gamma=3+\alpha~,~~\gamma x(s^\prime)=\alpha~. \]

Additionally, we obtain a differential relationship between $x$ and the averaged scale density at a given cut-off level (equation \ref{diff-relationship-3D} of which equation \ref{gen_x} is an exact solution). One could expect various application of this formula since it is derived for an arbitrary continuous pdf. 

 \item In the case of a power-law pdf with negative slope $q$ and its corresponding density profile $\rho(\ell) \propto \ell^{-p}$, we obtain scaling indices which are scale-free and functions of $q$:
\[ p=-\frac{3}{q},~~\gamma=3\frac{1+q}{q},~~\alpha=\frac{3}{q},~~x=\frac{1}{1+q} \] 
 
 \item Introducing a structure parameter $y$ in the 2D case to relate the averaged surface mass density and the mass of an abstract scale $\Sigma^\prime \propto (M^\prime)^y$ and under the assumption of point symmetry in the cloud, we obtain relations between the scaling indices in the 2D and 3D cases:
 \[ x=\frac{3y-1}{2}~,~~~\gamma=\frac{2}{1-y}~.\]
 Thus the 3D model is subject to observational test. Similar expressions are obtained by other authors while our contribution here is their derivation within the presented general framework. Using the relation between the 3D and 2D structure parameters $x$ and $y$, one could reconstruct from observational data the spatial structure of a class of equivalence.
 % Най-накрая искаме да подчертаем важността на разсъжденията в секция 5, които свързват нашия 3D модел с наблюденията, в термините на същата обща схема. Някои от математическите връзки в тази секция са получени от други автори, които сме цитирали съответно, но нашият принос е в общата постановка на това разглеждане. Чрез връзката между структурните параметри $x$ и $y$ (3D и 2D) ние можем да възстановим тримерната структура на класа на еквивалентност, ползвайки наблюдения (лесно може да се напише аналог на уравнение (21) за column density pdf). Друго важно приложение е получаването на скалиращите експоненти \gamma и \alpha, а също така построяването на CMF, ползвайки наблюдения.
 A differential relationship between the structure parameter $y$ and the averaged surface density at a given cut-off level is derived (equation \ref{diff-relationship-2D}), analogously to the 3D case.
 \item The proposed MC classes of equivalence as characterized by the scaling of the structure parameter $x$ are representative for the general structure of real clouds with various types of column-density pdfs: power law, lognormal or combination of both. In the case of power-law pdf, the predicted values of $x$ lead to mass functions of prestellar cores with slopes larger than the Salpeter value ($-1.35$) but close to it within the observational uncertainties.
\end{enumerate}

% , is fundamental for building one more sistematic theory of MCs. It might be as well to mention the differential relationship (\ref{diff-relationship-3D}) between averaged density at a given scale and structure parameter (of that scale) which allows us to calculate easy the structure parameter in the case of PL-tail.

\vspace{12pt}
{\it Acknowledgement:} T.V. acknowledges support by the {\em Deutsche Forschungsgemeinschaft} (DFG) under grant KL 1358/20-1. \\

% \newpage

\label{lastpage}

\end{document}